\documentstyle[onecolumn]{mn}

\title[Relativistic Approach to Gravitational Instability in the
Expanding Universe]
{A Relativistic Approach to Gravitational Instability in the
Expanding Universe: Second Order Lagrangian Solutions}

\author[S. Matarrese, O. Pantano and D. Saez]
{Sabino Matarrese$^{1}$,
Ornella Pantano$^{1}$,
and Diego Saez$^{2}$ \\
$^{1}$ Dipartimento di Fisica {\em Galileo Galilei}, Universit\`{a} di
Padova, via Marzolo 8, I--35131 Padova, Italy\\
$^{4}$ Departamento de Fisica Teorica, Universidad de Valencia,
Burjassot, Valencia, Spain\\}

\begin{document}

\maketitle

\begin{abstract}
A Lagrangian relativistic approach to the non--linear dynamics
of cosmological perturbations of an irrotational collisionless fluid
is considered. Solutions are given at second order in perturbation theory
for the relevant fluid and geometric quantities and compared with the
corresponding ones in the Newtonian approximation. Specifically, we
compute the density, the volume expansion scalar, the shear, the
``electric" part, or tide, and the ``magnetic" part of the Weyl tensor.
The evolution of the shear and the tide beyond the linear regime strongly
depends on the ratio of the characteristic size of the perturbation to
the cosmological horizon distance. For perturbations on sub--horizon scales
the usual Newtonian approximation applies, at least at the considered
perturbative order; on super--horizon scales, instead, a new picture emerges,
which we call ``silent universe", as each fluid element evolves
independently of the environment, being unable to exchange signals
with the surrounding matter through either sound waves or gravitational
radiation. For perturbations inside the Hubble radius particular attention is
paid in singling out non--local effects during the non--linear evolution of
fluid elements. These non--local effects are shown to be carried by a
traceless and divergenceless tensor, contained in the magnetic part
of the Weyl tensor, which is dynamically generated as soon as the system
evolves away from the linear regime.

\end{abstract}

\begin{keywords}
cosmology: theory -- gravitation -- galaxies: large--scale structure
of the Universe
 \end{keywords}

\section{Introduction}
The study of the dynamics of self--gravitating fluids is of extreme
importance in the study of the formation of structures in the
universe for from primordial perturbations via the gravitational instability.
The solutions of the relevant dynamical equations are obtained by different
techniques depending on the specific application. A Newtonian approach is
usually applied for perturbations on scales much smaller than the horizon
distance, but much larger than the Schwarzschild radii of collapsing objects
\cite{peebles80}. Analytical methods are then used to follow the evolution
of small perturbations and useful approximate solutions are also available to
describe the system up to the mildly non--linear regime, such as the
celebrated Zel'dovich approximation, while the general
non--linear dynamics can only be followed numerically, e.g. by N--body codes.
A relativistic treatment in cosmology is instead applied to study linear
perturbations on large scales and/or at early times, and to the analysis of
those few symmetric solutions of Einstein's equations which are relevant in a
cosmological context (e.g. the spherically symmetric Tolman--Bondi metric).

Having a unified treatment of the problem, able to describe both the
large--scale properties of the universe and the non--linear collapse of
structures by gravitational instability on small scales would be highly
desirable. Building up such a theory is however hampered
by the absence of symmetries in the system, which implies that all the
matter and geometric degrees of freedom are possibly switched on
during the non--linear evolution of self--gravitating matter.

A possible picture of the problem is represented by the
fluid--flow approach, in which the dynamics is followed
directly in terms of observable fluid variables (such as density,
shear, vorticity, etc.) and other tensor quantities which describe
the space--time curvature (instead of the metric tensor).
A complete description of the method can be found in the classical review
by Ellis (1971). Following this line of thought, Matarrese, Pantano \& Saez
(1993) have recently proposed a Lagrangian, general relativistic approach to
the fully non--linear dynamics of a collisionless fluid. The
approach is based on two relevant simplifying assumptions: vanishing
vorticity (i.e. irrotational flow) and negligible signal exchange by the
gravitational field, as it is obtained by
dropping the terms containing the so--called magnetic part $H_{ab}$ of the
Weyl tensor in the non--linear evolution of the tide (or electric
part of the Weyl tensor, $E_{ab}$). While the former assumption is a widely
accepted one in the standard gravitational instability scenario of structure
formation, the latter is certainly more problematic.
The meaning of $H_{ab}$ in linear theory is well known: it only contains
vector and tensor modes (e.g. Bruni, Dunsby \& Ellis 1992); if the vorticity
vanishes, as in our case, no physical vector modes are present and
$H_{ab}$ at most contains gravitational waves. Beyond linear theory,
however, the physical content of $H_{ab}$ is less clear. One can
reasonably assume that in the absence of $H_{ab}$ no gravitational
waves occur: this is evident in a Lagrangian approach, where,
the absence of pressure gradients and the vanishing of
$H_{ab}$ imply that no spatial
gradients appear in the evolution equations (apart from those possibly
contained in convective time derivatives, which can be dropped
by going to the comoving matter frame).
It is then clear that no actual wave propagation can occur when spatial
derivatives are absent in the fluid and gravitational evolution equations.
The method proposed by Matarrese, Pantano \& Saez (1993) takes advantage of
the absence of spatial gradients to construct a Lagrangian algorithm in which
the dynamics of fluid elements is completely determined by local initial
conditions; each fluid element can be therefore independently followed up
to the time when it enters a multi--stream region. In
this sense the method is quite
similar to the well--known Zel'dovich approximation in the Newtonian context,
to which it actually reduces to first order in a perturbation expansion.

While neglecting the fluid pressure (i.e. taking $p=0$) and considering zero
initial vorticity $\omega_{ab}$ are reasonable assumptions in a cosmological
framework, the condition $H_{ab}=0$ has to be taken with more caution.
It has been shown \cite{br89} that the only solutions of Einstein equations,
with $p=\omega_{ab}=H_{ab}=0$ are either of Petrov type I, or conformally
flat, or homogeneous and anisotropic of Bianchi type I, or with two degenerate
shear eigenvalues and described by a Szekeres line--element.
All of these cases require some restrictions on the initial data, which means
that the exact conditions above are not suitable for studying cosmological
structure formation with enough generality; requiring $p=\omega_{ab}=0$ and
$H_{ab} \approx 0$ appears more feasible.
In a recent work Matarrese, Pantano \& Saez (1994) have clarified
the dynamical role of the magnetic tensor, by calculating the behaviour of
fluctuations around Robertson--Walker (RW) at second order in perturbation
theory: whatever initially scalar perturbations are given, $H_{ab}$ vanishes
at first order, but not beyond.
The important consequence of this result is that the non--vanishing
$H_{ab}$ allows for the influence of the surrounding matter on the
evolution of fluid elements, through the presence of spatial gradients
in the evolution equations. The magnetic tensor appears therefore responsible
for the environmental influence of the matter on the non--linear evolution
of fluid elements. As we shall see below, although the signal carrying such a
non--local information travels at finite speed, for perturbations on scales
much smaller than the horizon distance it appears as an instantaneous Newtonian
effect.

In this paper we give a detailed derivation of the perturbative solutions at
second order of the fluid and gravitational field equations, expressed in
Lagrangian form. We pay particular attention to the occurrence of non--local
effects in the evolution of fluid elements away from the linear regime.
This work complements and extends the results previously reported in
(Matarrese, Pantano \& Saez 1994).

The plan of the paper is as follows. In Section 2 we review the general
fluid--flow method used to study the relativistic dynamics of a
self--gravitating pressureless fluid, which we then specialize to the
case of irrotational motions represented in a comoving frame, i.e. in
Lagrangian form; to this aim we introduce suitably rescaled variables which
describe deviations out of a homogeneous and isotropic universe expansion. In
Section 3 we present our second order expansion and discuss its properties
on scales larger and smaller than the horizon distance. Some cosmological
consequences of our work are discussed in Section 4. Two
appendices are added to help the
comparison with the Newtonian approximation and to provide some technical
details on the derivation of the second order expressions.

We use the signature $+2$; latin indices refer to space--time
coordinates, $(1,2,3,4)$, greek indices to spatial ones,
$(1,2,3)$; semicolons are used for covariant derivatives, commas for partial
ones. Units are such that the speed of light $c$ is one.

\section{Relativistic Dynamics}

In this Section we shall introduce the equations which govern the dynamics of
a collisionless perfect fluid in general relativity. The
dynamical equations are written in terms of observable fluid quantities
and other tensor quantities which directly
describe the space--time curvature. A complete description and a full
derivation of the equations presented here can be found in \cite{ellis71}.

\subsection{Relativistic hydrodynamical equations}

The relativistic dynamics of a collisionless
(i.e. with vanishing pressure) self--gravitating perfect fluid is determined
by Einstein's field equations and by the continuity
equations for the matter stress--energy tensor
$T_{ab} = \varrho u_a u_b$, where $\varrho$ is the energy density
(or matter density, in this case) and $u^a$ the four--velocity of the fluid,
normalized to $u^a u_a=-1$.
The local projection tensor into the rest frame of an observer moving
with four--velocity $u^a$ is defined as $h^{ab} \equiv g^{ab} + u^a u^b$,
for which $h_{ab}u^b=0$. By differentiating the velocity field one
obtains the space--like tensor $v_{ab} \equiv h_a^{~c} h_b^{~d} u_{c;d}$,
$v_{ab}u^b=0$, and the acceleration vector
$\dot u^a \equiv u^a_{~;b} u^b$, which is also space--like,
$\dot u^a u_a=0$, as it follows from the $u^a$ normalization. An overdot
denotes convective differentiation with respect to the proper time $t$ of
fluid elements (for a general $n$--index tensor,
${\dot A}_{a_1 a_2 \cdots a_n} = A_{a_1 a_2 \cdots a_n;b} u^b$).

Being flow--orthogonal, the tensor $v_{ab}$ can only have nine independent
components: its skewsymmetric is the vorticity tensor
$\omega_{ab} \equiv v_{[ab]}$ (the symbol $_{[..]}$
stands for skewsymmetrization, $_{(..)}$ for symmetrization),
which describes rigid rotations of fluid elements with respect to a locally
inertial rest frame. Due to its antisymmetric character, $\omega_{ab}$
has only three independent components which correspond to
the vorticity vector $\omega^a = {1 \over 2} \eta^{abcd}
u_b \omega_{cd}$, with $\eta_{abcd}$ the four--index Levi--Civita
tensor. Equivalently, $\omega_{ab}= \eta_{abcd} \omega^c u^d$.
The symmetric part of $v_{ab}$ can be further split into its trace
$\Theta \equiv v^a_{~a}$, called
the volume expansion scalar and its traceless part, $\sigma_{ab}
\equiv v_{(ab)} - {1 \over 3} \Theta h_{ab}$, called the shear
tensor. The volume expansion scalar, giving the local rate of isotropic
expansion (or contraction), defines a length--scale
$\ell$, through the equation $\Theta=3\dot \ell /\ell$. This
length--scale is just the generalization of the scale--factor $a(t)$
of homogeneous and isotropic RW models; in that particular case
$\Theta=3H$, where $H(t)$ is Hubble's constant.
The shear, on the other hand, describes a pure straining in which a
spherical fluid volume is distorted into an ellipsoid with axis lengths
changing at rates determined by the three $\sigma^a_{~b}$ eigenvalues,
$\sigma_1$, $\sigma_2$ and $\sigma_3=-(\sigma_1+\sigma_2)$.
The vanishing trace condition implies
that this deformation leaves the fluid volume invariant, while, in the absence
of vorticity, the principal axes of the shear keep their direction
fixed during the evolution, in a local inertial rest frame.

For a collisionless perfect fluid the energy--momentum local conservation
equations reduce to
\begin{equation}
\dot u^a = 0.
\end{equation}
\begin{equation}
{\dot \varrho} = - \varrho \Theta,
\end{equation}
Eq.(1) tells us that
in absence of pressure gradients each fluid element moves along a geodesic,
while Eq.(2) is equivalent to the continuity
equation for the particle number density, since the matter and energy
density are directly proportional in our case.

The expansion scalar satisfies the Raychaudhuri equation
\begin{equation}
{\dot \Theta} = \Lambda - {1 \over 3} \Theta^2 + 2 (\omega^2 - \sigma^2)
- 4 \pi G\varrho,
\end{equation}
where $G$ is Newton's constant and we have introduced a possible
cosmological constant $\Lambda$ and the two scalars
$\omega^2 \equiv \omega^a \omega_a = {1 \over 2} \omega^{ab} \omega_{ab}$
and $\sigma^2 \equiv {1 \over 2} \sigma^{ab} \sigma_{ab}$.
Note that in the homogeneous and isotropic RW case, $\omega_{ab}=\sigma_{ab}=0$
and the latter equation reduces to the familiar Friedmann equation,
$3(\dot H + H^2) = - 4 \pi G \varrho + \Lambda$.

The vorticity vector evolves according to
\begin{equation}
{\dot \omega}^a = - {2 \over 3} \Theta \omega^a + \sigma^a_{~b} \omega^b ,
\end{equation}
while the shear is determined by the equation
\begin{equation}
{\dot \sigma}_{ab} = - \sigma_{ac} \sigma^c_{~b} - \omega_a \omega_b +
{1 \over 3} h_{ab} (2 \sigma^2 + \omega^2)
- {2 \over 3} \Theta \sigma_{ab} - E_{ab},
\end{equation}
where $E_{ac}\equiv C_{abcd}u^bu^d$ is the electric part of
the Weyl tensor $C_{abcd}$. The tensor $E_{ab}$ is also
called tidal tensor, since it contains that component of the
gravitational field which describes tidal interactions; it is symmetric,
traceless and flow--orthogonal, $E_{ab}u^b=0$.

{}From the Weyl tensor one can define another quantity: its magnetic part
$H_{ac}={1 \over 2} \eta_{ab}^{~~~gh}C_{ghcd}u^bu^d$
which is also symmetric, traceless and flow--orthogonal.
While the tidal field has a straightforward Newtonian analogue, which
can be written in terms of derivatives of the gravitational potential,
the magnetic tensor $H_{ab}$ has no straightforward interpretation in the
Newtonian equations. The important point is that, while in Newtonian theory,
in its standard formulation, the gravitational
potential is determined through a constraint, namely
Poisson's equation, in general relativity both $E_{ab}$ and $H_{ab}$ can
be calculated by solving suitable evolution equations.

{}From the Bianchi identities one can derive the following evolution equations
for $E_{ab}$ and $H_{ab}$
\begin{equation}
{\dot E}_{ab}=
- h_{c(a} h_{b)m} \eta^{mrsd} u_r H^c_{~s;d}
- h_{ab} \sigma^{cd} E_{cd}- \Theta E_{ab} +
E_{c(a} \omega_{b)}^{~~c} +
3 E_{c(a} \sigma_{b)}^{~~c} - 4 \pi G \varrho \sigma_{ab},
\end{equation}
\begin{equation}
{\dot H}_{ab}=
 h_{c(a} h_{b)m} \eta^{mrsd} u_r E^c_{~s;d}
- h_{ab} \sigma^{cd} H_{cd}- \Theta H_{ab} +
H_{c(a} \omega_{b)}^{~~c} +
3 H_{c(a} \sigma_{b)}^{~~c}.
\end{equation}

In a collisionless fluid, the velocity field satisfies the following
constraint equations
\begin{equation}
\omega^a_{~;a} = 0.
\end{equation}
\begin{equation}
h^a_b (\omega^{bc}_{~~;c} - \sigma^{bc}_{~~;c}  +
{2 \over 3} \Theta^{;b}) = 0,
\end{equation}
Further constraints need to be satisfied by the electric and the
magnetic tensor, namely
\begin{equation}
H_{ad} = h_a^{~t} h_d^{~s} (\omega_{(t}^{~b;c} + \sigma_{(t}^{~b;c})
\eta_{s)fbc} u^f
\end{equation}
\begin{equation}
h^t_{~a} E^{as}_{~~;d} h_s^{~d} - \eta^{tbpq} u_b \sigma_p^{~d} H_{qd}
+ 3 H^t_{~s} \omega^s = {8 \pi G \over 3} h^t_{~b} \rho^{;b}
\end{equation}
\begin{equation}
h^t_{~a} H^{as}_{~~;d} h_s^{~d} - \eta^{tbpq} u_b \sigma_p^{~d} E_{qd}
+ 3 E^t_{~s} \omega^s = \rho \omega^t
\end{equation}

Finally, another useful quantity is the (infinitesimal) relative
position vector $x_{\perp}^{~a}$ between two neighbouring particles
defined in the rest frame of the observer moving with four--velocity
$u^{\alpha}$ ($x_{\perp a} u^a =0$). The vector components $x_{\perp}^{~a}$
evolves according to the ``generalized Hubble law"
\begin{equation}
\dot x_{\perp}^{~a} = {1 \over 3} \Theta ~x_{\perp}^{~a}  +
(\omega^a_{~b} + \sigma^a_{~b} ) ~x_{\perp}^{~b} .
\end{equation}

\subsection{Cosmological equations}

In a cosmological context one can safely assume zero initial
vorticity for the fluid, i.e. $\omega^a=0$; Kelvin's circulation
theorem implies that this condition continues to hold until the fluid enters
a multi--stream region. Under the assumption of vanishing magnetic tensor,
Matarrese, Pantano \& Saez (1993) proposed a method for integrating the
evolution equations in local inertial frames, taking advantage of the
geodesics motion of collisionless particles. However, when spatial gradients
explicitly occur, as it happens when $H_{ab} \neq 0$, it is more
convenient to adopt a global coordinate system, such as one comoving with
the fluid. Let us then rewrite the equations of Section 2.1 in comoving
coordinates. Recalling that for a
pressureless fluid comoving hypersurfaces (orthogonal to the energy flow) and
synchronous ones (orthogonal to geodesics) coincide, one can write the
line--element in the form
\begin{equation}
ds^2= - dt^2+a^2(t) {\tilde h}_{\alpha\beta} dq^\alpha dq^\beta,
\end{equation}
where $a = A t^{2/3}$, as for a flat, matter--dominated RW
model with vanishing cosmological constant (our ``background" solution) and
the rescaled spatial projection tensor
${\tilde h}_{\alpha\beta}$ generally depends upon space and time coordinates.
For computational convenience we then introduce suitably rescaled quantities
\cite{mps94}: a
scaled density fluctuation $\Delta \equiv (6\pi G t^2 \varrho -1)/a$,
a peculiar expansion scalar $\vartheta = (3t/2a)(\Theta - 2/t)$, a
traceless shear tensor $s^\alpha_{~\beta} \equiv (3t/2a)
\sigma^\alpha_{~\beta}$ and a traceless tidal tensor
$e^\alpha_{~\beta} \equiv (3t^2/2a) E^\alpha_{~\beta}$.
These quantities can be grouped in two space--like tensors:
the velocity gradient tensor $\vartheta^\alpha_{~\beta} \equiv
s^\alpha_{~\beta} + {1\over 3}
\delta^\alpha_{~\beta} \vartheta$, related to the
covariant derivatives of the peculiar velocity field and
the peculiar gravitational field tensor
$\Delta^\alpha_{~\beta} \equiv e^\alpha_{~\beta} + {1 \over 3}
\Delta \delta^\alpha_{~\beta}$. We also rescale the magnetic tensor
as ${\cal H}^\alpha_{~\beta} \equiv (3t^2/2a) H^\alpha_{~\beta}$.
The dynamical equations for the fluid and the gravitational field then read
\begin{equation}
{\vartheta}^{\alpha~\prime}_{~\beta} =
-\frac{3}{2a} (\vartheta^\alpha_{~\beta}
+ \Delta^\alpha_{~\beta}) - \vartheta^\alpha_{~\gamma}
\vartheta^\gamma_{~\beta} \ ,
\end{equation}

\begin{eqnarray}
{\Delta}^{\alpha~\prime}_{~\beta} =
 -  \frac{1}{a} (\vartheta^\alpha_{~\beta}
+ \Delta^\alpha_{~\beta})
- 2(\vartheta \Delta^\alpha_{~\beta} + \Delta \vartheta^\alpha_{~\beta}) +
\frac{5}{2} \Delta^\alpha_{~\gamma} \vartheta^\gamma_{~\beta}
+ \frac{1}{2} \Delta^\gamma_{~\beta}\vartheta^\alpha_{~\gamma} +
\delta^\alpha_{~\beta}(\Delta \vartheta -
\Delta^\gamma_{~\delta}\vartheta^\delta_{~\gamma})
\nonumber \\
+ \frac{3 t}{4 a^2}
{\tilde h}_{\beta\eta} \bigl( {\tilde \eta}^{\eta\gamma\delta}
{\cal H}^\alpha_{~\gamma};_{\delta} + {\tilde \eta}^{\alpha\gamma\delta}
{\cal H}^\eta_{~\gamma};_{\delta} \bigr)\ ,
\end{eqnarray}

\begin{equation}
{\cal H}^{\alpha~\prime}_{~\beta} =
- \frac{1}{a} {\cal H}^\alpha_{~\beta}
- 2 \vartheta {\cal H}^\alpha_{~\beta} - \delta^\alpha_{~\beta}
\vartheta^\gamma_{~\delta} {\cal H}^\delta_{~\gamma} +
\frac{5}{2} {\cal H}^\alpha_{~\gamma} \vartheta^\gamma_{~\beta}
+\frac{1}{2} {\cal H}^\gamma_{~\beta}\vartheta^\alpha_{~\gamma}
- \frac{3 t}{4 a^2} {\tilde h}_{\beta\eta}
\bigl({\tilde \eta}^{\eta\gamma\delta}
\Delta^\alpha_{~\gamma};_{\delta} + {\tilde \eta}^{\alpha\gamma\delta}
\Delta^\eta_{~\gamma};_{\delta} \bigr)\ ,
\end{equation}
where the prime denotes partial differentiation with respect to the scale
factor $a$ and ${\tilde \eta}^{\alpha\gamma\delta}$ is the
Levi--Civita tensor relative to the conformal spatial metric
${\tilde h}_{\alpha\beta}$, namely ${\tilde \eta}^{\alpha\beta\gamma} =
{\tilde h}^{-1/2} \varepsilon^{\alpha\beta\gamma}$, with $\varepsilon^{123}=1$.
The traces of Eqs.(15) and (16) yield, respectively,
\begin{equation}
{\vartheta}^{\prime} =
-\frac{3}{2a} (\vartheta + \Delta) - \vartheta^\alpha_{~\gamma}
\vartheta^\gamma_{~\alpha} ,
\end{equation}
\begin{equation}
{\Delta}^{\prime} = - \vartheta \Delta
 -  \frac{1}{a} (\vartheta + \Delta) ,
\end{equation}
while the trace of Eq.(17) vanishes identically. As shown by Matarrese,
Pantano \& Saez (1994), Eq.(15) coincides with the equation
obtained by taking the spatial gradients of the Newtonian Euler equation in
Lagrangian form; Eq.(19) is instead identical to the Newtonian continuity
equation, also in Lagrangian form (see Appendix A1). Since the traceless
component of the evolution equation for $\Delta^\alpha_{~\beta}$ (i.e. the
evolution equation for the tide) has no Newtonian analog, it
follows that the magnetic tensor which enters through its spatial gradients
in that equation, has no direct dynamical role in the Newtonian theory as it
is usually formulated (i.e. in the standard form of Appendix A1).

In our comoving gauge, the spatial metric tensor, which appears explicitly and
through its spatial gradients in the covariant derivatives of Eqs.(16) and
(17), evolves according to the equation
\begin{equation}
{1 \over 2} {\tilde h}^{\alpha\gamma} {{\tilde h}^{\prime}}_{\gamma\beta} =
\vartheta^\alpha_{~\beta} .
\end{equation}
Using now our rescaled variables we can rewrite the constraints Eqs.(9)--(12)
as
\begin{equation}
{\vartheta_\alpha^{~\beta}};_\beta = \vartheta,_\alpha  ,
\end{equation}
\begin{equation}
{\cal H}_{\alpha}^{~\beta} = { t \over 2 a} {\tilde h}_{\alpha\mu} \bigl(
{\tilde \eta}^{\mu\gamma\delta} \vartheta_\gamma^{~\beta};_\delta +
{\tilde \eta}^{\beta\gamma\delta} \vartheta_\gamma^{~\mu};_\delta \bigr) ,
\end{equation}
\begin{equation}
{\Delta_\alpha^{~\beta}};_\beta = \Delta,_\alpha - \frac{2 a^2}{3 t}
{\tilde h}_{\alpha\mu} {\tilde h}_{\beta\nu} {\tilde \eta}^{\mu\lambda\gamma}
\vartheta^\nu_{~\lambda} {\cal H}^\beta_{~\gamma} ,
\end{equation}
\begin{equation}
{\cal H}_\alpha^{~\beta};_\beta = \frac{2 a^2}{3 t} {\tilde h}_{\alpha\mu}
{\tilde h}_{\beta\nu} {\tilde \eta}^{\mu\lambda\gamma}
\vartheta^\nu_{~\lambda} \Delta^\beta_{~\gamma} .
\end{equation}

Note that the only Christoffel symbols involved in the covariant derivatives
of Eqs.(16), (17) and (21) -- (24) are the spatial ones, i.e.
$\Gamma^\alpha_{\beta\gamma} = {\tilde \Gamma}^\alpha_{\beta\gamma}$, which
can be trivially obtained from ${\tilde h}_{\alpha\beta}$.

The rescaled displacement vector $\xi^{\alpha} = x_{\perp}^{~\alpha}/ a$
obeys the equation
\begin{equation}
\xi^{\alpha\, \prime} = \vartheta^\alpha_{~\beta} \xi^\beta .
\end{equation}
The matrix connecting the Eulerian coordinates $x^\alpha$ with the
Lagrangian ones $q^\beta$ is the Jacobian
\begin{equation}
J^\alpha_{~\beta} \equiv \partial x^\alpha / \partial q^\beta \equiv
\delta^\alpha_{~\beta} + {\cal D}^\alpha_{~\beta} ,
\end{equation}
where ${\cal D}^\alpha_{~\beta}$ is the symmetric deformation tensor.
Taking then $\xi^\alpha = d x^\alpha = J^\alpha_{~\beta}
\xi_{(0)}^{~\beta}$, where $\xi_{(0)}^{~\beta} = dq^\beta$ represents the
initial (i.e. Lagrangian) infinitesimal displacement, one gets
\begin{equation}
{\cal D}^{\alpha~\prime}_{~\beta}  = \vartheta^\alpha_{~\beta} +
\vartheta^\alpha_{~\gamma}{\cal D}^\gamma_{~\beta},
\end{equation}
which is formally solved by
\begin{equation}
{\cal D}^\alpha_{~\beta}(a) = \exp\int_{a_0}^a
d \overline a ~\vartheta^\alpha_{~\beta}(\overline a )
- \delta^\alpha_{~\beta}.
\end{equation}

Once the Jacobian is known one gets the metric at the ``time'' $a$ by
performing the transformation
\begin{equation}
{\tilde h}_{\alpha\beta}(a) = {\tilde h}_{\gamma\delta}(a_0) J^\gamma_{~\alpha}
J^\delta_{~\beta}.
\end{equation}

\subsection{Initial conditions}
Initial perturbations can be given at the ``time" $a_0$ in a RW background
by using the gauge--invariant formalism (Bardeen 1980). Here we consider only
growing--mode scalar perturbations \cite{mps93},
\begin{equation}
\Delta^\alpha_{~\beta} (a_0) =
- \vartheta^\alpha_{~\beta}(a_0) = \varphi_0,^\alpha_{~\beta},
\end{equation}
where the scalar $\varphi_0$, an arbitrary function of the spatial
coordinates $q^\alpha$, corresponds to the initial peculiar gravitational
potential, which is related to Bardeen's gauge--invariant $\Phi_H$
by $\varphi_0 = - (3 / 2 A^3) \Phi_H$.
These initial conditions correspond to the ``seed" metric
\begin{equation}
{\tilde h}_{\alpha\beta} = \delta_{\alpha\beta}(1 - \frac{20}{9}
A^3 \varphi_0) - 2 a \varphi_0,_{\alpha\beta},
\end{equation}
and imply vanishing magnetic tensor ${\cal H}^\alpha_{~\beta}$ at the linear
level. In what follows, we will only consider the growing mode, proportional
to $a\varphi_0,_{\alpha\beta}$, since the constant mode,
proportional to $A^3 \varphi_0 \ll 1$, can be practically neglected during the
subsequent evolution. Also, in many cases it will be convenient to assume
that the initial conditions are given at $a_0 \to 0$.

It is important to realize that the constraints, Eqs.(21)--(24), are all
satisfied at the linear level \cite{bde92} by our choice of initial
conditions.

\section{Second Order Solutions: Beyond the Zel'dovich Approximation}

We shall now construct a second order Lagrangian perturbation expansion
in the amplitude of the fluctuations around the RW background solution. To
this aim, it will prove useful to define the two scalar quantities
\cite{doros70}
\begin{equation}
\mu_1 \equiv \varphi_0,^\gamma_{~\gamma} = \lambda_1 +
\lambda_2 + \lambda_3
\end{equation}
\begin{equation}
\mu_2 \equiv  {1 \over 2}(\varphi_0,^\gamma_{~\gamma}
\varphi_0,^\delta_{~\delta} - \varphi_0,^\gamma_{~\delta}
\varphi_0,^\delta_{~\gamma}) =
\lambda_1\lambda_2 + \lambda_1\lambda_3 + \lambda_2\lambda_3,
\end{equation}
where $\lambda_\alpha$ are the local eigenvalues of the symmetric tensor
$\varphi_0,^\alpha_{~\beta}$.
Let us then expand the relevant quantities retaining only terms
up to second order,
\begin{equation}
\vartheta^\alpha_{~\beta}
= \vartheta^{~~\alpha}_{(1)\beta} + \vartheta^{~~\alpha}_{(2)\beta}
\end{equation}
\begin{equation}
\Delta^\alpha_{~\beta}
= \Delta^{~~\alpha}_{(1)\beta} + \Delta^{~~\alpha}_{(2)\beta}
\end{equation}
\begin{equation}
{\cal H}^\alpha_{~\beta}
=  {\cal H}^{~~\alpha}_{(2)\beta}
\end{equation}
where the zeroth order terms of $\vartheta^\alpha_{~\beta}$ and
$\Delta^\alpha_{~\beta}$ vanish, while the expansion of
${\cal H}^\alpha_{~\beta}$ starts from second order.

For the velocity gradient tensor $\vartheta^\alpha_{~\beta}$
and the peculiar gravitational field tensor $\Delta^\alpha_{~\beta}$ the
first order terms obviously coincide with the initial data,
\begin{equation}
\Delta^{~~\alpha}_{(1)\beta} =
- \vartheta^{~~\alpha}_{(1)\beta} = \varphi_0,^\alpha_{~\beta} ,
\end{equation}
while the second order corrections read (for more details see Appendix A2)
\begin{equation}
\vartheta^{~~\alpha}_{(2)\beta} =
 {a \over 7} \bigl( - 12 \mu_1 \varphi_0,^\alpha_{~\beta}
 + 6 \mu_2 \delta^\alpha_{~\beta} + 5
 \varphi_0,^\alpha_{~\gamma} \varphi_0,^\gamma_{~\beta} \bigr)
 + \chi^\alpha_{~\beta} ,
\end{equation}
\begin{equation}
\Delta^{~~\alpha}_{(2)\beta} =
 {a \over 7} \bigl( 20 \mu_1 \varphi_0,^\alpha_{~\beta}
 - 10 \mu_2 \delta^\alpha_{~\beta}
 - 13 \varphi_0,^\alpha_{~\gamma} \varphi_0,^\gamma_{~\beta} \bigr)
 + \kappa^\alpha_{~\beta} .
\end{equation}
Here indices are raised by the Kronecker symbol.
The traceless tensors $\chi^\alpha_{~\beta}$ and $\kappa^\alpha_{~\beta}$
represent the contributions coming from the magnetic part of the Weyl tensor to
$\vartheta^\alpha_{~\beta}$ and $\Delta^\alpha_{~\beta}$, respectively,
and have zero divergence like the magnetic tensor ${\cal H}^\alpha_{~\beta}$
itself at this order (see Eqs.(A12)--(A13)).

{}From Eqs.(37)--(39) one immediately obtains the traces
\begin{equation}
\vartheta = - \mu_1 + a(- \mu_1^2 + {8 \over 7} \mu_2)
\end{equation}
and
\begin{equation}
\Delta = \mu_1 + a(\mu_1^2 - {4 \over 7} \mu_2) ,
\end{equation}
which coincide with those obtained in the Lagrangian second order Newtonian
theory \cite{buchert89,buchert92,mabpr91,bouchet92,gramann93,lr93}.

We can perform a second--order expansion also for the spatial displacement
vector $\xi^\alpha$
\begin{equation}
\xi^\alpha = \xi^{~~\alpha}_{(0)} + \xi^{~~\alpha}_{(1)} +
\xi^{~~\alpha}_{(2)}
\end{equation}
{}From Eqs.(25) and (37) one immediately gets
\begin{equation}
\xi^{~~\alpha}_{(1)} = - a \varphi_0,^\alpha_{~\beta} \xi^{~~\alpha}_{(0)}
\end{equation}
\begin{equation}
\xi^{~~\alpha}_{(2)} = - a \varphi_0,^\alpha_{~\beta} \xi^{~~\beta}_{(1)}
+ \int_{a_0}^a d \overline a \vartheta^{~~\alpha}_{(2)\beta}
{}~\xi^{~~\beta}_{(0)} \equiv a^2 \
\psi^{\alpha}_{~\beta} ~\xi^{~~\beta}_{(0)} ,
\end{equation}
where $\psi^\alpha_{~\beta}$ is given by
\begin{equation}
\psi^\alpha_{~\beta} = {3 \over 7} \bigl( - 2 \mu_1
\varphi_0,^\alpha_{~\beta} + \mu_2 \delta^\alpha_{~\beta} + 2
\varphi_0,^\alpha_{~\gamma} \varphi_0,^\gamma_{~\beta}\bigr) + {1 \over a^2}
\int_{a_0}^a d \overline a ~\chi^\alpha_{~\beta} ,
\end{equation}
with trace $\psi^\alpha_{~\alpha} = - {3 \over 7} \mu_2$.
Since $dx^\alpha \equiv \xi^\alpha$ and $dq^\alpha \equiv \xi^{~\alpha}_{(0)}$
one can write
\begin{equation}
dx^\alpha = dq^\alpha - a \varphi_0,^\alpha_{~\beta} dq^\beta +
a^2 \psi^\alpha_{~\beta} d q^\beta.
\end{equation}
Then, the second order perturbative solution for the deformation tensor reads
\begin{equation}
{\cal D}^\alpha_{~\beta} \equiv - a \varphi_0,^\alpha_{~\beta} + a^2
\psi^\alpha_{~\beta} ;
\end{equation}
we can immediately recognize that the first order term is nothing but the
kinematical Zel'dovich approximation \cite{zel70} while the second order
correction is provided by the symmetric tensor $\psi^\alpha_{~\beta}$.

In Newtonian theory one would write the same formal expression, but the
irrotationality condition would automatically lead to $\psi^\alpha_{~\beta} =
\psi,^\alpha_{~\beta}$, with the potential $\psi$ satisfying the second order
Poisson equation
\begin{equation}
\nabla^2 \psi = - {3 \over 7} \mu_2 ,
\end{equation}
which is consistent with the trace of the relativistic equation (45).
This implies that the Newtonian eigenvalues $\nu_\alpha$ of
$\psi^\alpha_{~\beta}$ only need to satisfy the
condition $\sum_\alpha \nu_\alpha = - {3 \over 7} \mu_2$, while
in order to get the complete information on the single
$\nu_\alpha$'s one needs the Newtonian definition of
$\psi^\alpha_{~\beta}$ as $\psi,^\alpha_{~\beta}$, i.e. a non--local
information. The relativistic $\nu_\alpha$'s also solve the Newtonian
equations, but the reverse is not necessarily true: it just depends upon the
specific boundary conditions used in solving Poisson's equation.

Finally, from Eqs.(26), (29) we can compute the conformal spatial metric
tensor
\begin{equation}
{\tilde h}_{\alpha\beta} = \delta_{\alpha\beta}
- 2 a \varphi_0,_{\alpha\beta} +
{a^2 \over 7} \bigl(19 \varphi_0,_{\alpha\gamma} \varphi_0,^\gamma_{~\beta}
- 12 \mu_1 \varphi_0,_{\alpha\beta} +
6 \mu_2 \delta_{\alpha\beta} \bigr) +
\int_{a_0}^a d \overline a \chi_{\alpha\beta} \ .
\end{equation}

The traceless, divergenceless symmetric tensor $\chi^\alpha_{~\beta}$,
representing the contribution to the velocity gradient tensor caused by the
magnetic part
of the Weyl tensor, can be written as a convolution
\begin{equation}
\chi^\alpha_{~\beta}({\bf q},a) = \int d^3 q'
S^\alpha_{~\beta} ({\bf q'})
f(\vert {\bf q} - {\bf q'}\vert,a)
\end{equation}
of the source tensor
\begin{equation}
{\cal S}^\alpha_{~\beta} =
{\mu_2},^\alpha_{~\beta} - \nabla^2 (2 \mu_1
\varphi_0,^\alpha_{~\beta} - 2 \varphi_0,^\alpha_{~\gamma}
\varphi_0,^\gamma_{~\beta} - \delta^\alpha_{~\beta} \mu_2 ) ,
\end{equation}
with the time--dependent function $f$, whose Fourier transform $\hat f(k,a)$
satisfies the third order linear differential equation
\begin{equation}
{d^3 \over dx^3}\hat f +
{9 \over x} {d^2 \over dx^2}\hat f +
\left( {12 \over x^2} + 1 \right)  {d \over dx}\hat f
+ {3 \over x} \hat f
= C x ,
\end{equation}
where $x = k\tau$, with $\tau = (3/A) t^{1/3}$ the conformal time, and
$C \equiv 10 A^3 / 21 k^4$.
The initial conditions are $\hat f(x_0) = {d \over dx }\hat f(x_0)
= {d^2 \over d x^2}\hat f(x_0) = 0$.

\subsection{Fate of cosmological perturbations}

Let us now analyse the evolution of the solution of Eq.(52),
describing the behaviour of perturbations on scales much larger ($x \gg 1$)
and much smaller ($x \ll 1$) than the horizon scale.

\subsubsection{Outside the horizon: the silent universe}

In the limit $x \ll 1$ Eq.(51) reduces to
\begin{equation}
{d^3 \over dx^3}\hat f +
{9 \over x} {d^2 \over dx^2}\hat f +
 {12 \over x^2}   {d \over dx}\hat f
\approx  C x ;
\end{equation}
setting $\hat g= \hat f + {C \over 180} x^4$ Eq.(53) reads
\begin{equation}
{d^3 \over dx^3}\hat g +
{9 \over x} {d^2 \over dx^2}\hat g +
{12 \over x^2}   {d \over dx}\hat g
\approx  0 ,
\end{equation}
whose formal solutions are $\hat g=\hat g_0 x^n$ with $n=-1,-5$.
Keeping only the growing mode, we then find
\begin{equation}
\hat f \approx {C \over 180} x^4 = {1 \over 378} A^3 \tau^4 .
\end{equation}
When $x\ll 1$, i.e. when the characteristic scale of the perturbation exceeds
the horizon distance, $\chi^\alpha_{~\beta} \approx
(3 t^2 / 14 a) {\cal S}^\alpha_{~\beta}$, and the  contribution to
$\vartheta^\alpha_{~\beta}$ due to the magnetic tensor becomes negligible;
a similar reasoning would apply to $\Delta^{\alpha}_{~\beta}$.
The relevant expressions can then be obtained from Eqs.(38), (39) and (50):
\begin{equation}
\vartheta^{\alpha}_{~\beta} \approx - \varphi_0,^\alpha_{~\beta} +
{a \over 7} \bigl( - 12 \mu_1 \varphi_0,^\alpha_{~\beta}
+ 6 \mu_2 \delta^\alpha_{~\beta} + 5
\varphi_0,^\alpha_{~\gamma} \varphi_0,^\gamma_{~\beta} \bigr) ,
\end{equation}
\begin{equation}
\Delta^{\alpha}_{~\beta} \approx \varphi_0,^\alpha_{~\beta} +
{a \over 7} \bigl( 20 \mu_1 \varphi_0,^\alpha_{~\beta}
- 10 \mu_2 \delta^\alpha_{~\beta}
- 13 \varphi_0,^\alpha_{~\gamma} \varphi_0,^\gamma_{~\beta} \bigr) ,
\end{equation}
\begin{equation}
{\cal H}^{\alpha}_{~\beta}
\approx {6 t \over 7}
\bigl(\varepsilon^\alpha_{~~\gamma\delta} \varphi_0,^\delta_{~\beta}
\varphi_0,^\gamma_{~\nu} + \varepsilon_{\beta\gamma\delta}
\varphi_0,^{\delta\alpha} \varphi_0,^\gamma_{~\nu}\bigr),^\nu.
\end{equation}
Perturbations with size larger than the Hubble radius evolve as a separate
silent universe, with spatial gradients playing no role. Note that the
expression in Eq.(55), and the related results in Eqs.(56) -- (58) could have
been obtained by taking the $c \to 0$ limit in Eq.(52).

Unfortunately, these general relativistic effects have little implications
for the problem of cosmological structure formation, since perturbations on
super--horizon scales are usually assumed to have very small amplitude, so
that a linear approximation is sufficient. Conversely, the validity of the
silent universe approximation on ultra--large scales might suggest its use as
a mathematical tool to look for cosmologically interesting background solutions
of Einstein's field equations. Nevertheless, there are a number of formal
consequences of our relativistic solutions, which is worth mentioning.
One of these is the absence of two--dimensional configurations.
This can be seen as follows: if one eigenvalue of $\varphi_0,^\alpha_{~\beta}$,
e.g. $\lambda_3$, vanishes everywhere, the Newtonian theory, with suitable
boundary conditions, would imply $\vartheta_3(a)=0$ or $x_3(a)=q_3$, i.e. no
motion along the third axis. This is usually referred as ``two--dimensional"
gravitational clustering.
As far as the second order deformation tensor is concerned, one would have
$\nu_1+\nu_2 = - {3 \over 7} \mu_2$, with $\mu_2=\lambda_1\lambda_2$, and
$\nu_3=0$.
In the general relativistic case with ${\cal H}^{\alpha}_{~\beta}$, instead,
we immediately find $\nu_1 = \nu_2 = - \nu_3 = - {3 \over 7} \mu_2$,
and $\vartheta_3(a) \neq 0$ for $a \neq a_0$.
The motion dynamically impressed along the third axis soon becomes of the same
order of magnitude as that in the other two directions. This effect
is to be ascribed to the tide--shear coupling term
$\delta^\alpha_{~\beta} (\vartheta \Delta - \Delta^\gamma_{~\delta}
\vartheta^\delta_{~\gamma})$ in the evolution equation for the tide,
which reduces to $- 2 \mu_2 \delta^\alpha_{~\beta}$ to lowest perturbative
order, namely second order.
The only case when this coupling disappears is when two $\lambda_\alpha$'s
simultaneously vanish, i.e. for planar symmetry. Therefore $\vartheta_3(a)=0$
is not an exact solution of the relativistic equations, unless another
$\vartheta_\alpha$ also vanishes. As an example,
no axisymmetric configurations without motion along the symmetry axis are
allowed.

\subsubsection{Inside the horizon: validity of the Newtonian approximation}

In the limit $x \gg 1$ it is convenient to define the function $\hat g$
as follows
\begin{equation}
\hat g \equiv {d \over dx} \hat f + {3 \over x} \hat f
\end{equation}
where $\hat g$ now obeys the equation
\begin{equation}
{d^2 \over dx^2}\hat g + \hat g
\approx  Cx ,
\end{equation}
with initial conditions at $x_{\star} \sim 1$ equal to
$\hat g({x_\star}) \approx A^3 x_{\star}^3/54 k^4$ and
$d{\hat g}/dx({x_\star}) \approx A^3 x_{\star}^2/18 k^4$ .
The general solution of the above equation reads:
\begin{equation}
\hat g(x) = \hat g_1(x) + \hat g_2(x) + C \left(
\hat g_1 \int^x_{x_{\star}} {\hat g_2 \over W} \, \bar x d \bar x
- \hat g_2 \int^x_{x_{\star}} {\hat g_1 \over W} \, \bar x d \bar x \right),
\end{equation}
where $\hat g_1$ and $\hat g_2$ are independent solutions of the
homogeneous wave equation ${d^2 \over dx^2}\hat g + \hat g=0$
and $W={d\hat g_1 \over dx}\hat g_2 -\hat g_1{d\hat g_2 \over dx} $
is the Wronskian. Since
\begin{equation}
\hat g_1 =b_1 \, \cos x, \qquad \hat g_2=b_2 \, \sin x, \qquad
W= -b_1 b_2,
\end{equation}
with $b_1$, $b_2$, integration constants, the general solution reads
\begin{equation}
\hat g =b_1 \, \cos x + b_2 \, \sin x +C \left(
x-x_{\star} \cos (x-x_{\star}) - \sin (x-x_{\star}) \right).
\end{equation}
Keeping only the growing mode, we find
\begin{equation}
{d \over dx} \hat f + {3 \over x} \hat f \approx Cx
\quad \longrightarrow \quad
\hat f \approx {C x^2 \over 5} = {2 A^3 \tau^2  \over 21 k^2}
\end{equation}
If $k\tau \gg 1$ we find
\begin{equation}
\chi^\alpha_{~\beta} \approx {6a \over 7}
\left( 2 \mu_1 \varphi_0,^\alpha_{~\beta} -
2 \varphi_0,^\alpha_{~\gamma} \varphi_0,^\gamma_{~\beta}
-\delta^\alpha_{~\beta} \mu_2 \right)
+ 2 a \psi,^\alpha_{~\beta}
\end{equation}
and
\begin{equation}
\kappa^\alpha_{~\beta} \approx - {5 \over 3} \chi^\alpha_{~\beta}.
\end{equation}
The second order deformation tensor reduces to $\psi^\alpha_{~\beta}=
\psi,^\alpha_{~\beta}$, while the conformal spatial metric tensor reads
\begin{equation}
{\tilde h}_{\alpha\beta} = \delta_{\alpha\beta}
- 2 a \varphi_0,_{\alpha\beta} + a^2 \psi,_{\alpha\beta} .
\end{equation}
The remaining relevant quantities reduce to
\begin{equation}
\vartheta^\alpha_{~\beta} \approx - \varphi_0,^\alpha_{~\beta} -
a \varphi_0,^\alpha_{~\gamma} \varphi_0,^\gamma_{~\beta} +
2 a \psi,^\alpha_{~\beta},
\end{equation}
\begin{equation}
\Delta^\alpha_{~\beta} \approx \varphi_0,^\alpha_{~\beta}
+a \left(\varphi_0,^\alpha_{~\gamma} \varphi_0,^\gamma_{~\beta}
- {10 \over 3} \psi,^\alpha_{~\beta}
\right).
\end{equation}
All these expressions coincide with those obtained at second
order in the Newtonian approximation in Lagrangian form
\cite{buchert89,buchert92,mabpr91,bouchet92,gramann93,lr93}
and can be obtained
from the $c \to \infty$ limit of Eq.(52). The scalar $\psi$ carries
information on the influence of the surrounding matter on the
non--linear dynamics of fluid elements.
Note that $\psi,^\alpha_{~\beta}$ produces a tilt of the principal
axes of the first--order deformation tensor, $\varphi_0,^\alpha_{~\beta}$.

For the magnetic tensor we find
\begin{equation}
\dot {\cal H}^\alpha_{~\beta} +
{\dot a \over a}{\cal H}^\alpha_{~\beta} \approx 0 .
\end{equation}
Then, within the horizon ${\cal H}^\alpha_{~\beta}$ decays as
\begin{equation}
{\cal H}^\alpha_{~\beta}(a) \approx {a_H  \over a}
{\cal H}^\alpha_{~\beta}(a_H) ,
\end{equation}
where the subscript $H$ refers to the moment when the considered wavelength
crosses the Hubble radius.

A numerical integration of Eq.(52) shows the range of validity of
the approximate solutions in Eqs.(55) and (64); this is shown in Figure 1,
where we plot the numerical solution of Eq.(52) vs. the two asymptotic
expressions.

\begin{figure}
\vspace{8cm}

\caption{Behaviour of the function $\hat f$ vs. $x=k\tau$; the dashed lines
refer to the two asymptotic expressions in Eqs.(55) and (64), respectively
for $x \ll 1$ and $x \gg 1$.}
\end{figure}

\subsection{Solutions with zero magnetic tensor}

Let us consider now the evolution of an infinite homogeneous ellipsoid,
for which ${\cal H}^\alpha_{~\beta}$ (this can be seen by noting that
the source tensor in Eq.(51) has no Fourier modes inside the Hubble radius).
For this discussion it is more convenient to consider as dynamical variables
$\vartheta$, $\Delta$, the shear tensor
$s^\alpha_{~\beta} = \vartheta^\alpha_{~\beta} - {1 \over 3}
\delta^\alpha_{~\beta} \vartheta$ and the tidal one $e^\alpha_{~\beta} =
\Delta^\alpha_{~\beta} - {1 \over 3} \delta^\alpha_{~\beta} \Delta$.
For vanishing magnetic tensor, the shear, the tidal field and the metric can
be simultaneously diagonalized.
The only  non--vanishing components of $s^\alpha_{~\beta}$ and
$e^\alpha_{~\beta}$ are the eigenvalues $s_\alpha$ and $e_\alpha$.
In terms of the eigenvalues $\lambda_\alpha$ of the tensor
$\varphi^{~~\alpha}_{0,~\beta}$ we have
\begin{equation}
s_\alpha= - \lambda_\alpha + {1 \over 3} \mu_1 +
a \bigl( -{12 \over 7} \mu_1 \lambda_\alpha + {5 \over 7} \lambda^2_\alpha +
{1 \over 3} \mu^2_1 + {10 \over 21} \mu_2 \bigr)
\end{equation}
\begin{equation}
e_\alpha=  \lambda_\alpha - {1 \over 3} \mu_1 +
a \bigl( {20 \over 7} \mu_1 \lambda_\alpha - {13 \over 7} \lambda^2_\alpha -
{1 \over 3} \mu^2_1 - {26 \over 21} \mu_2 \bigr) ,
\end{equation}
while the expressions for $\vartheta$ and $\Delta$ are given by Eqs.(40) and
(41).

An alternative expression for the density can be obtained
by replacing the volume expansion scalar in the exact continuity equation.
This will be called the ``continuity density" and denoted by a subscript
$c$. We easily find
\begin{equation}
1 + a \Delta_c = (1 + a_0 \mu_1) \prod_{\alpha=1}^3 (1 - a \lambda_\alpha +
a^2 \nu_\alpha)^{-1} ,
\end{equation}
with
\begin{equation}
\nu_\alpha = {3 \over 7} \mu_2 - { 6 \over 7} \lambda_\alpha
(\lambda_{\alpha - 1} + \lambda_{\alpha + 1})
\end{equation}
the eigenvalues of $\psi^\alpha_{~\beta}$ (the notation being such that
if $\alpha=1$, $\alpha-1=3$, while, if
$\alpha=3$, $\alpha+1=1$). Note that $\Delta_c$ only coincides with $\Delta$
up to second order, i.e. $\Delta_c=\Delta + {\cal O}(a^2)$.

Let us now give the second order solution for some
configurations with degenerate shear and tide eigenvalues, i.e. $s_1=s_2$ and
$e_1=e_2$. For simplicity, we set $a_0=0$ in what follows.

\smallskip
\noindent
{\it Sphere} --
In this case $\lambda=\lambda_1=\lambda_2=\lambda_3$, and the eigenvalues
of the shear and tide vanish identically, while
\begin{equation}
\vartheta= - 3 \lambda \bigl( 1 +  {13 \over 7} a \lambda \bigr)
\end{equation}
and
\begin{equation}
\Delta = 3 \lambda \bigl( 1 +  {17 \over 7} a \lambda \bigr) ,
\end{equation}
\begin{equation}
1 + a \Delta_c =
\left(1 - a \lambda - {3 \over 7} a^2 \lambda^2 \right)^{-3} .
\end{equation}

\smallskip
\noindent
{\it Pancake} --
In this case $\lambda_1=\lambda_2=0$, then $s_1=s_2$  and $e_1=e_2$.
The second order solution is
\begin{equation}
s_1 = - e_1 = {\lambda_3 \over 3} ( 1 +  a \lambda_3 ) ,
\end{equation}
\begin{equation}
\vartheta = - \Delta = - \lambda_3 ( 1 +  a \lambda_3 ) ,
\end{equation}
while
\begin{equation}
1 + a \Delta_c = ( 1 -  a \lambda_3 )^{-1} ,
\end{equation}
coincides with the exact comoving density (cfr. Matarrese, Pantano \& Saez
1993).

\smallskip
\noindent
{\it Filament} --
In this case we have $\lambda_1=\lambda_2=\lambda$ and $\lambda_3=0$.
Again, the shear and the tidal tensors have two equal eigenvalues
which at second order read
\begin{equation}
s_1=s_2 = - {\lambda \over 3} \bigl( 1 - {19 \over 7} a \lambda \bigr) ,
\end{equation}
\begin{equation}
e_1=e_2 = {\lambda \over 3} \bigl(  1 - {27 \over 7} a \lambda \bigr) ,
\end{equation}
while
\begin{equation}
\vartheta = -2 \lambda \bigl(  1 - {10 \over 7} a \lambda \bigr)
\end{equation}
and
\begin{equation}
\Delta = 2 \lambda \bigl(  1 + {12 \over 7} a \lambda \bigr) ,
\end{equation}
\begin{equation}
1 + a \Delta_c = \left(1 - a \lambda - {3 \over 7} a^2 \lambda^2 \right)^{-2}
\left(1 + {3 \over 7} a^2 \lambda^2 \right)^{-1} .
\end{equation}

\section{Discussion}

Our second order perturbative approach to the general relativistic dynamics
of irrotational dust can be used to provide some insight on the main dynamical
issue: the fate of general perturbations. So far, only a few analytical
solutions of the Lagrangian system of Eqs.(15) -- (24) are known: for exact
planar symmetry, $\lambda_1=\lambda_2=0$, one recovers the Zel'dovich pancake
solution (Zel'dovich 1970), as shown by Matarrese, Pantano \& Saez (1993),
which, in a relativistic context can be seen as a particular Szekeres model
(Szekeres 1975; see also the discussion by Kasai 1992, 1993 and Croudace et
al. 1994); in the
spherically symmetric case, $\lambda_1=\lambda_2=\lambda_3$, the system admits
the so--called Tolman--Bondi solution (Tolman 1934, Bondi 1947;
see also the discussion by Matarrese, Pantano \& Saez 1993), which, from a
local point of view, has a direct Newtonian analog in the well--known
top--hat model (e.g. Peebles 1980).
A systematic study of the local system of equations (i.e. that obtained
by setting the magnetic tensor to zero) has been carried out by Croudace et al.
(1994), who looked for solutions representing attractors
among general trajectories. They found that both spherical collapse and a
perfect pancake are repellers for general initial conditions, but
argued that the pancake instability would probably disappear
in the presence of a non--zero ${\cal H}^\alpha_{~\beta}$.

Bertschinger \& Jain (1994) have shown that the instability of the
pancake solution in the local system of equations is caused
by the tide--shear coupling in the evolution of the tide, which has a
destabilizing effect on the pancake solution (for general initial conditions)
but stabilizes prolate configurations. According to this analysis, for
vanishing ${\cal H}^\alpha_{~\beta}$, a strongly prolate spindle with
expansion along its axis is the generic outcome of collapse, except for
specific initial conditions corresponding to exactly spherical or planar
configurations. They proposed a Newtonian interpretation for this collapse
theorem and argued that, even though in the
most general case ${\cal H}^\alpha_{~\beta}$ is likely to be non--zero,
it can be set to zero for highly symmetrical configurations and probably
neglected in many other circumstances.

As stressed by Matarrese, Pantano \& Saez (1994),
the tide--shear coupling term in the tide evolution equation
is not present in the equations of Newtonian theory, at least in its standard
formulation, although it is clearly compatible with them.
The dominant role of this term for the non--linear fluid evolution is a
peculiarity of the purely local equations, which certainly applies to
perturbations on super--horizon scales, but cannot be easily understood in
the Newtonian limit. As a further illustration of this interpretation
let us consider the collapse of an infinite homogeneous ellipsoid,
which is described by our equations when ${\cal H}^\alpha_{~\beta}=0$.
In such a case the Newtonian dynamics is known to lead to the formation of
oblate spheroids (e.g. White \& Silk 1979; Barrow \& Silk 1981),
pancake--like objects with one collapsing axis and the other two tending to
a finite size (apart from initial conditions corresponding to an initially
prolate spheroid). The general relativistic dynamics, instead,
favours the formation of prolate objects, collapsing filaments with
expansion along their axis. This point is discussed in some detail
by Zel'dovich \& Novikov (1983), who argue that the discrepant behaviour
is due to the different role of boundary conditions at infinity in the
two theories. Our second order calculations show that the evolution
of fluid elements as isolated ellipsoids cannot apply to perturbations on
scales smaller than the Hubble radius, where non--local effects
would play a fundamental role. The appearance of non--local terms in general
solutions of the Newtonian theory has been recently stressed by Kofman (1994).

Given that the magnetic part of the Weyl tensor is generated during the
mildly non--linear evolution of irrotational dust, even in case it was
primordially set to zero, is there any application left for a purely local
treatment on cosmologically relevant scales? One can try to make some guess.
It may be that during the late phases of the evolution of a given fluid
element, i.e. in the strongly non--linear regime, the role of the magnetic
tensor becomes negligible again, possibly due to the fact that some sort
of (approximate) local axisymmetry is dynamically established. However, if this
picture should prove correct one cannot assume that the principal axes of the
collapsing ellipsoid coincide with the initial ones; in fact, already
at second order in perturbation theory the effect of the magnetic tensor is
that of tilting these axes, on account of the interaction with the
environment. A further complication is induced by the occurrence
of orbit crossing with the subsequent formation of multi--streaming.

Let us finally mention what we think is an important point.
The main lesson we have learned from the purely local treatment,
our silent universe approach, is that one could try to isolate
two competing effects: i) If ${\cal H}^\alpha_{~\beta}$ is switched off,
the fate of the collapse of a general fluid element is completely determined by
its local initial conditions; in such a case, one would generally expect
prolate rather than oblate collapse.
ii) In the real world, however, ${\cal H}^\alpha_{~\beta}$ is non--zero
except for a number of unrealistic cases listed in the introduction;
${\cal H}^\alpha_{~\beta}$ carries information on the action of the
``rest of the world" on the considered fluid element, i.e. what can be called
the effects of the environment.
In other words, the matter surrounding the fluid element acts by
compressing or stretching, and generally deforming it, thus modifying its local
dynamics in a way which cannot be easily predicted a priori. The relevant
issue is: which of these two competing effects is going to dominate
on the long run? It seems to us that the answer generally depends
upon various variables: the scenario (e.g. cold or hot dark matter, amount of
baryons, etc.), and the overall initial conditions, as they are
usually implied by the power--spectrum and the statistics of
the primordial density fluctuation field.

\smallskip
While completing this work two preprints have circulated (Bertschinger \&
Hamilton 1994; Kofman \& Pogosyan 1994), where the non--linear evolution
equation for the electric component of the Weyl tensor is derived from
Newtonian gravity.
The presence of a non--vanishing non--local contribution in this limit agrees
with our previous results (Matarrese, Pantano \& Saez 1994)
on the behaviour of the general relativistic equations inside the Hubble
radius.

\section* {Acknowledgments}
M. Bruni is acknowledged for many useful discussions.
This work has been partially supported by Italian MURST. DS thanks the
Conselleria de Cultura, Educacio i Ciencia de la Generalitat Valenciana
and the Spanish DGICYT project PB90--0416 for financial support.

\appendix
\section{Newtonian dynamics}

The equations which govern the non--linear dynamics of a collisionless fluid
in Newtonian theory for an expanding universe can be written as
(e.g. Peebles 1980)
\begin{equation}
{d \delta \over dt} + (1+\delta) \nabla \cdot {\bf v} = 0,
\end{equation}
\begin{equation}
{d {\bf v} \over dt} + 2 {{\dot a} \over a} {\bf v}
+ { 1 \over a^2} \nabla \phi = 0,
\end{equation}
\begin{equation}
\nabla^2 \phi = {3 \over 2} {\dot a}^2 \delta .
\end{equation}
where ${\bf v} \equiv d {\bf x} / d t$ is the peculiar velocity, $\phi$ the
peculiar gravitational potential and $\delta \equiv (6 \pi G t^2 \rho -1)$ the
fractional density contrast.
In order to compare these equations with the relativistic ones it is convenient
to use rescaled variables for the velocity field
${\bf u} \equiv d{\bf x}/da = {\bf v}/{\dot a}$, the density fluctuation field
$\Delta = \delta /a$ and the local gravitational potential
$\varphi \equiv (3/2A^3)\phi$ and to adopt as time variable the scale factor
$a$ itself (e.g. Gurbatov, Saichev \& Shandarin 1989; Matarrese et al. 1992).
Eqs.(A1--A3) in the new variables read
\begin{equation}
\Delta' +
u^\beta \Delta,_\beta = - {1 \over a} (u^\beta,_\beta + \Delta) -
\Delta u^\beta,_\beta ,
\end{equation}
\begin{equation}
{u^\alpha}' + u^\beta u^\alpha,_\beta = - {3 \over 2 a} (u^\alpha +
\varphi,^\alpha) ,
\end{equation}
\begin{equation}
\varphi,^\beta_{~\beta} = \Delta .
\end{equation}
By differentiating the
Euler equation (A5), defining the symmetric tensors
$\vartheta^\alpha_{~\beta} \equiv u^\alpha,_\beta$,
$\Delta^\alpha_{~\beta} \equiv \varphi,^\alpha_{~\beta}$, and
adopting a Lagrangian description, one recovers Eq.(15),
while the continuity equation (A4) coincides with the trace of Eq.(16).
It is clear that the Newtonian approximation described above is degenerate,
as it provides only one equation to fully determine the symmetric tensor
$\Delta^\alpha_{~\beta}$: any traceless tensor added to the r.h.s. of Eq.(16),
leaves the Newtonian equations unchanged. In order to
completely determine the evolution of the gravitational field tensor
$\Delta^\alpha_{~\beta}$ one has to resort to its definition
in terms of the potential $\varphi$, i.e. to a non--local
theory. Because of the intrinsic non--locality of the Newtonian approximation
one needs boundary conditions to
determine the dynamics, contrary to the general relativistic equations,
where Cauchy data on a spatial hypersurface are enough;
this comes from the fact that the
Poisson equation (A6) is an elliptic, constraint equation.
It is well known (e.g. Ellis 1971) that the
lack of evolution equations for the traceless part of the gravitational
tensor, $e^\alpha_{~\beta}$, implies that the
Newtonian theory may add spurious solutions which would be
discarded by the full relativistic system.

\section{Second Order Solutions}

At second order the differential equations (15)--(17) become
(for convenience the integration variable is now the proper time of fluid
elements)
\begin{equation}
\dot\vartheta^{~~\alpha}_{(2)\beta}
+ {3 \dot a \over 2a} \left( \vartheta^{~~\alpha}_{(2)\beta}
+\Delta^{~~\alpha}_{(2)\beta} \right)
+ \dot a \varphi_0,^\alpha_{~\gamma}\varphi_0,^\gamma_{~\beta}
= 0 ,
\end{equation}
\begin{equation}
\dot \Delta^{~~\alpha}_{(2)\beta}
+ {\dot a \over a} \left( \vartheta^{~~\alpha}_{(2)\beta}
+\Delta^{~~\alpha}_{(2)\beta} \right)
+ \dot a \left(3 \varphi_0,^\alpha_{~\gamma}\varphi_0,^\gamma_{~\beta}
-4\mu_1 \varphi_0,^\alpha_{~\beta} +2 \mu_2 \delta^\alpha_{~\beta} \right)
=
{1 \over 2a }\left(
\epsilon_{\beta}^{~\gamma \delta} {\cal H}^\alpha_{~\gamma},_{\delta}
+ \epsilon_{~\gamma}^{\alpha ~ \delta} {\cal H}^\gamma_{~\beta},_{\delta}
\right) ,
\end{equation}
\begin{eqnarray}
\dot {\cal H}^{\alpha}_{~\beta} + {\dot a \over a}{\cal H}^{\alpha}_{~\beta}
= - {1 \over 2 a} \left(
\epsilon_{\beta}^{~\gamma \delta} \Delta^{~~\alpha}_{(2)\gamma},_{\delta}
+ \epsilon_{~\gamma}^{\alpha ~ \delta} \Delta^{~~\gamma}_{(2)\beta},_{\delta}
\right)
+  {1 \over 2 } \left(
\epsilon_{\beta}^{~\gamma \delta} \varphi_0,^{\alpha}_{~\mu \delta}
+ \epsilon^{\alpha \gamma \delta} \varphi_0,_{\beta \mu \delta}
\right) \varphi_0,^{\mu}_{~ \gamma}
\nonumber \\
+ {5 A^3 \over 3 a} \left(
\epsilon_{\beta}^{~\gamma \delta} \varphi_0,^{\alpha}_{~\gamma}
\epsilon^{\alpha \gamma \delta} \varphi_0,_{\beta \gamma}
\right) \varphi_0,_{\delta} ,
\end{eqnarray}
where the last term in Eq.(A9) has been presented for completeness, but
it is a decaying mode and will be neglected in what follows. Note that
in order to write Eqs.(A8) and (A9) we used the Christoffel
symbols of the three--metric ${\tilde h}_{\alpha\beta}$ evaluated at first
order; these are trivially obtainable in terms of the initial
gravitational potential $\varphi_0$ and its derivatives.

In the expressions for $\vartheta^{~~\alpha}_{(2)\beta}$
and $\Delta^{~~\alpha}_{(2)\beta}$ it is convenient to isolate those
parts (indicated by a tilde) which solve the system (A7)--(A9) with
vanishing magnetic term ${\cal H}^{\alpha}_{~\beta}$. It is possible to
proceed in this way because the contributions from the magnetic term start
only from second order. We have
\begin{equation}
\vartheta^{~~\alpha}_{(2)\beta}={\tilde \vartheta}^{~~\alpha}_{(2)\beta}
 + \chi^\alpha_{~\beta} ,
\end{equation}
\begin{equation}
\Delta^{~~\alpha}_{(2)\beta}={\tilde \Delta}^{~~\alpha}_{(2)\beta}
 + \kappa^\alpha_{~\beta} .
\end{equation}
The traceless tensors $\chi^\alpha_{~\beta}$ and $\kappa^\alpha_{~\beta}$
represent the contributions caused by the presence of the magnetic part
of the Weyl tensor and have zero divergence,
\begin{equation}
\chi^{~\alpha}_{\beta~ ;\alpha} = \kappa^{~\alpha}_{\beta~ ;\alpha} = 0 .
\end{equation}
Similarly, at second order, we have
\begin{equation}
{\cal H}^{~\alpha}_{\beta~;\alpha} = 0 .
\end{equation}
The previous conditions are directly derived from the constraint equations,
and are obviously consistent with the evolution equations.

For zero magnetic term, the solutions of Eqs.(A7) and (A8) can be easily
found; they read
\begin{equation}
{\tilde \vartheta}^{~~\alpha}_{(2)\beta} =
 {a \over 7} \bigl( - 12 \mu_1 \varphi_0,^\alpha_{~\beta}
 + 6 \mu_2 \delta^\alpha_{~\beta} + 5
 \varphi_0,^\alpha_{~\gamma} \varphi_0,^\gamma_{~\beta} \bigr)  ,
\end{equation}
\begin{equation}
{\tilde \Delta}^{~~\alpha}_{(2)\beta} =
 {a \over 7} \bigl( 20 \mu_1 \varphi_0,^\alpha_{~\beta}
 - 10 \mu_2 \delta^\alpha_{~\beta}
 - 13 \varphi_0,^\alpha_{~\gamma} \varphi_0,^\gamma_{~\beta} \bigr)  .
\end{equation}

The second order terms due to the magnetic tensor satisfy the following
equations
\begin{equation}
{\dot \chi}^{\alpha}_{~\beta}
+ {3 \dot a \over 2a} \left( \chi^{\alpha}_{~\beta}
+\kappa^{\alpha}_{~\beta} \right) = 0 ,
\end{equation}
\begin{equation}
{\dot \kappa}^{\alpha}_{~\beta}
+ {\dot a \over a} \left( \chi^{\alpha}_{~\beta}
+\kappa^{\alpha}_{~\beta} \right)
- {1 \over 2a}\left(
\epsilon_{\beta}^{~\gamma \delta} {\cal H}^\alpha_{~\gamma},_{\delta}
+ \epsilon^{\alpha \gamma \delta} {\cal H}_{\beta \gamma},_{\delta}
\right) = 0 ,
\end{equation}
\begin{equation}
{\dot {\cal H}}^{\alpha}_{~\beta} +{\dot a \over a}{\cal H}^{\alpha}_{~\beta}
= - {1 \over 2 a } \left(
\epsilon_{\beta}^{~\gamma \delta} \kappa^{\alpha}_{~\gamma},_{\delta}
+ \epsilon^{~\alpha \gamma \delta} \kappa_{\beta \gamma},_{\delta}
\right)
- {10 \over 7} \left( C^{\alpha}_{~\beta} + C_{\beta}^{~\alpha} \right) ,
\end{equation}
where
\begin{equation}
C^{\alpha}_{~\beta} \equiv
\epsilon^{\alpha \gamma \delta} \left(
\varphi_0,_{~\beta \gamma} \, \varphi_0,_{\delta \nu} \right),^\nu .
\end{equation}
One can easily eliminate ${\cal H}^{\alpha}_{~\beta}$ by differentiating
Eq.(A17) with respect to $t$. One gets
\begin{eqnarray}
{\ddot \kappa}^\alpha_{~\beta} + {2 \over t} {\dot \kappa}^\alpha_{~\beta}
+ {2 \over 3t} {\dot \chi}^\alpha_{~\beta}
+ {2 \over 9t^2} \left(
\chi^{\alpha}_{~\beta} +\kappa^{\alpha}_{~\beta}
\right)
= - {1 \over 4a^2}
\left[
2\epsilon^{\alpha \gamma \eta}
\epsilon^{~\mu \delta}_{\beta} \kappa_{\gamma \mu, \delta \eta}
+\epsilon^{~\mu \delta}_{\gamma}
\left( \epsilon^{~\gamma \eta}_{\beta} \kappa^{\alpha}_{~\mu, \delta \eta}
+\epsilon^{\alpha \gamma \eta} \kappa_{\beta \mu, \delta \eta}
\right) \right] \nonumber\\
- {5 \over 7 a}
\left[
\epsilon^{~\gamma \delta}_{\beta}
\left(
C^{\alpha}_{~\gamma, \delta} + C^{~\alpha}_{\gamma~,\delta}
\right)
+\epsilon^{\alpha \gamma \delta}
\left(
C_{\gamma \beta, \delta} + C_{\beta \gamma,\delta}
\right) \right] .
\end{eqnarray}
Using the relation
\begin{equation}
\epsilon^{\alpha \beta \gamma} \epsilon_{\mu \nu \sigma} =
3! \, \delta^{[\alpha}_{~\mu} \delta^{\beta}_{~\nu} \delta^{\gamma]}_{~\sigma}
,
\end{equation}
we then obtain
\begin{equation}
{\ddot \kappa}^\alpha_{~\beta} + {2 \over t} {\dot \kappa}^\alpha_{~\beta}
- {4 \over 9t^2} \left( \chi^{\alpha}_{~\beta}
+\kappa^{\alpha}_{~\beta} \right)
-{1 \over a^2} \nabla^2 \kappa^\alpha_{~\beta}
= - {10 \over 7a} S^\alpha_{~\beta},
\end{equation}
where the constant source tensor $S^\alpha_{~\beta}$ is given by
\begin{equation}
{\cal S}^\alpha_{~\beta} =
{\mu_2},^\alpha_{~\beta} - \nabla^2 (2 \mu_1
\varphi_0,^\alpha_{~\beta} - 2 \varphi_0,^\alpha_{~\gamma}
\varphi_0,^\gamma_{~\beta} - \delta^\alpha_{~\beta} \mu_2 ) .
\end{equation}
Replacing now
\begin{equation}
\kappa^{\alpha}_{~\beta}= - {d \over d t} \left(t \chi^{\alpha}_{~\beta}
\right)
\end{equation}
into the differential equation (A22) we get
\begin{equation}
{d^3 \over dt^3} \chi^\alpha_{~\beta}
+ {5 \over t} {d^2 \over dt^2} \chi^\alpha_{~\beta}
+ {32 \over 9t^2} {d \over dt} \chi^\alpha_{~\beta}
-{1 \over a^2} \nabla^2 \left( {d \over dt} \chi^\alpha_{~\beta}
-{1 \over t} \chi^\alpha_{~\beta} \right)
= - {10 \over 7at} S^\alpha_{~\beta},
\end{equation}
which has to be solved together with the initial conditions
\begin{equation}
\chi^\alpha_{~\beta} (t_0) = {d \over dt} \chi^\alpha_{~\beta} (t_0)
= {d^2 \over dt^2} \chi^\alpha_{~\beta} (t_0) = 0 ,
\end{equation}
since our linear initial data imply
$\chi^\alpha_{~\beta} (a_0)$ $ = \kappa^\alpha_{~\beta} (a_0)$
$= {\cal H}^\alpha_{~\beta} (a_0) =0$.

The traceless tensor $\chi^\alpha_{~\beta}$, representing the
contribution due to the magnetic part of the Weyl tensor, can be written
as a convolution
\begin{equation}
\chi^\alpha_{~\beta}({\bf q},a) = \int d^3 q'
S^\alpha_{~\beta} ({\bf q'})
f(\vert {\bf q} - {\bf q'}\vert,a)
\end{equation}
of the source $S^\alpha_{~\beta}$ with the function $f$, whose
Fourier transform $\hat f(k,t)$ satisfies the equation
\begin{equation}
{d^3 \over dt^3}\hat f +
{5 \over t} {d^2 \over dt^2}\hat f +
{32 \over 9t^2} {d \over dt}\hat f + {k^2 \over a^2}
\biggl( {d \over dt}\hat f + {1 \over t} \hat f \biggr)
= {10 \over 7at} ,
\end{equation}
with initial conditions
$\hat f(t_0) = \left[ d \hat f / dt \right] (t_0)
= \left[ d^2 \hat f/ dt^2 \right] (t_0) = 0$.
Note that the possibility of 3D Fourier expanding the previous quantities
is ensured by the underlying flat RW background.

In order to discuss the general behaviour of the previous equation it can be
useful to adopt as independent variable $x \equiv k \tau$ where
$\tau=(3/A) t^{1/3}$ is the conformal time. One then gets Eq.(52) whose
properties are discussed in the main text.
\end{document}